\documentclass[aps,prc,superscriptaddress,showpacs,floatfix,nofootinbib,notitlepage,twocolumn]{revtex4-1}
\usepackage{amsmath,graphicx,float,hyperref,upgreek}

\begin{document}

\title{Non-Gaussian transverse momentum fluctuations from impact parameter fluctuations}
\author{Rupam Samanta}
\affiliation{AGH University of Science and Technology, Faculty of Physics and
Applied Computer Science, aleja Mickiewicza 30, 30-059 Cracow, Poland}
\affiliation{Universit\'e Paris Saclay, CNRS, CEA, Institut de physique th\'eorique, 91191 Gif-sur-Yvette, France}
\author{Jo\~ao Paulo Picchetti}
\affiliation{Instituto de F\'{\i}sica, Universidade de  S\~{a}o Paulo,  Rua  do  Mat\~{a}o, 1371,  Butant\~{a},  05508-090,  S\~{a}o  Paulo,  Brazil}
\author{Matthew Luzum}
\affiliation{Instituto de F\'{\i}sica, Universidade de  S\~{a}o Paulo,  Rua  do  Mat\~{a}o, 1371,  Butant\~{a},  05508-090,  S\~{a}o  Paulo,  Brazil}
\author{Jean-Yves Ollitrault}
\affiliation{Universit\'e Paris Saclay, CNRS, CEA, Institut de physique th\'eorique, 91191 Gif-sur-Yvette, France}

\begin{abstract} 
The transverse momentum per particle, $[p_t]$, fluctuates event by event in ultrarelativistic nucleus-nucleus collisions, for a given multiplicity. 
These fluctuations are small and approximately Gaussian, but a non-zero skewness has been predicted on the basis of hydrodynamic calculations, and seen experimentally. 
We argue that the mechanism driving the skewness is that, if the system thermalizes, the mean transverse momentum increases with impact parameter for a fixed collision multiplicity.  
We postulate that fluctuations are Gaussian at fixed impact parameter, and that non-Gaussianities solely result from impact parameter fluctuations. 
Using recent data on the variance of $[p_t]$ fluctuations, we make quantitative predictions for their skewness and kurtosis as a function of the collision multiplicity.
We predict in particular a spectacular increase of the skewness below the knee of the multiplicity distribution, followed by a fast decrease.
\end{abstract}

\maketitle

There is now wide consensus that collisions between atomic nuclei at ultrarelativistic energies produce a tiny droplet of fluid made of quarks and gluons, which quickly thermalizes as a result of the strong interaction.
For two decades, evidence for the formation of a fluid has largely relied on the observation of anisotropic flow, seen through azimuthal correlations between outgoing particles~\cite{Heinz:2013th}. 
Evidence of a different nature has recently been revealed~\cite{Samanta:2023amp}, based on the fluctuations of the transverse momentum per particle, $[p_t]$, across collision events with the same multiplicity. 
These are traditional observables of nucleus-nucleus collisions~\cite{NA49:1999inh,CERES:2003sap,STAR:2003cbv,PHENIX:2003ccl,STAR:2005vxr,ALICE:2014gvd}, which are used to constrain theoretical models~\cite{Broniowski:2009fm,Bernhard:2019bmu,JETSCAPE:2020shq,Nijs:2020ors}.
The new observation by the ATLAS collaboration is that the variance of these fluctuations in Pb+Pb collisions decreases by a factor $\sim 2$ over a narrow multiplicity range~\cite{ATLAS:2022dov} corresponding to ultracentral collisions~\cite{Luzum:2012wu,CMS:2013bza,Plumari:2015cfa,Shen:2015qta,Carzon:2020xwp,Liu:2022kvz,Giannini:2022bkn,Kuroki:2023ebq}, where $[p_t]$ fluctuations had not yet been analyzed. 
This decrease is naturally explained by invoking thermalization. 
Thermalization indeed implies that $[p_t]$ is linked with the density, which depends on impact parameter.
The observed decrease then results from the decrease of impact parameter fluctuations in collisions with the largest multiplicity.

In this paper, we show that this mechanism also implies that fluctuations of $[p_t]$  are strongly non-Gaussian in ultracentral collisions.\footnote{It has already been observed that hydrodynamic calculations imply a significant skewness of $[p_t]$ fluctuations~\cite{Giacalone:2020lbm}, but the crucial role of impact parameter has so far been overlooked.} 
Using the same model of $[p_t]$ fluctuations as in Ref.~\cite{Samanta:2023amp}, we make quantitative, parameter-free predictions for the skewness and excess kurtosis, which are standard measures of the non-Gaussianity. 

\begin{figure}[h]
\begin{center}
\includegraphics[width=0.9\linewidth]{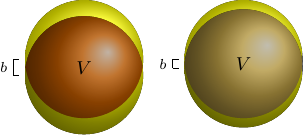} 
\end{center}
\caption{
  Representation of Pb+Pb collisions with the same multiplicity and different impact parameters $b=1.8$~fm (left) and $b=1.0$~fm (right), corresponding to centrality fractions $c_b\simeq 1.5$~\% and $c_b\simeq 0.5$~\%. 
  The interval between these two values is the typical spread of $c_b$ for fixed multiplicity. 
  The larger value of $b$ goes along with a smaller collision volume, implying a larger density, symbolically depicted with a darker color. 
}
\label{fig:cartoon}
\end{figure} 
In order to understand the fluctuations of $[p_t]$ at fixed multiplicity $N_{ch}$, one must take into account two effects.
The first is that collision events with the same multiplicity $N_{ch}$ can have different impact parameter $b$. 
This simple fact is often overlooked because $N_{ch}$ is traditionally used as an estimator of the centrality, as defined by $b$.
The second effect is that for a given multiplicity $N_{ch}$, $[p_t]$ depends on impact parameter $b$. 
Larger $b$ implies a smaller collision volume $V$ (Fig.~\ref{fig:cartoon}), hence larger density $N_{ch}/V$.
If the system thermalizes, the temperature is higher and the momentum per particle $[p_t]$ is larger.

We first explain the origin of non-Gaussian fluctuations on the basis of a simplified model, where $[p_t]$ is a single-valued function of $N_{ch}$ and $b$.
Instead of $b$, we use the centrality fraction $c_b\simeq \pi b^2/\sigma_{\rm Pb}$~\cite{Das:2017ned} (where $\sigma_{\rm Pb}$ is the inelastic cross section of the Pb+Pb collision) as an equivalent variable throughout this paper. 
The variation of $c_b$ for fixed $N_{ch}$ is small enough that the dependence of $[p_t]$ on $c_b$ can be linearized:\footnote{Note that observables depend quadratically on $b$ for small $b$ for symmetry reasons~\cite{Pepin:2022jsd}, which forbids a dependence of the type $\sqrt{c_b}$ for small $b$.}
\begin{equation}
  \label{linearized}
[p_t]=p_t^{\rm min}+\lambda c_b,
\end{equation}
where $p_t^{\rm min}$ and $\lambda$ depend on $N_{ch}$.

The probability distribution of $[p_t]$ is then determined by that of $c_b$. 
The probability distribution of $c_b$ for fixed $N_{ch}$, $p(c_b|N_{ch})$, can easily be determined~\cite{Das:2017ned}. 
First, one assumes that the distribution of $N_{ch}$ at fixed impact parameter is Gaussian:
\begin{equation}
\label{gaussianN}
  p(N_{ch}|c_b)=\frac{1}{\sqrt{2\pi}\sigma_{N_{ch}}(c_b)}\exp\left(-\frac{\left(N_{ch}-\overline{N_{ch}}(c_b)\right)^2}{2\sigma_{N_{ch}}(c_b)^2}\right). 
\end{equation}
For ultracentral collisions where $c_b\ll 1$, one neglects the dependence of $\sigma_{N_{ch}}$ on $c_b$, and linearizes the variation of the mean:
\begin{equation}
  \label{linearizedmean}
\overline{N_{ch}}(c_b)=N_{\rm knee}-\beta c_b,
\end{equation}
where $N_{\rm knee}$ is the knee, defined as the average multiplicity for $b=0$, and $\beta$ determines the decrease of the multiplicity with centrality. 
The values of these parameters can be obtained by fitting the measured distribution of $N_{ch}$~\cite{Das:2017ned}.
In our numerical calculations, we use the values appropriate for Pb+Pb collisions at $\sqrt{s_{NN}}=5.02$~TeV, and for the charged particle multiplicity seen by the inner detector of ATLAS, namely: $N_{\rm knee}=3680$, $\sigma_{N_{ch}}=168$, $\beta=18300$~\cite{Samanta:2023amp}.

\begin{figure}[h]
\begin{center}
\includegraphics[width=\linewidth]{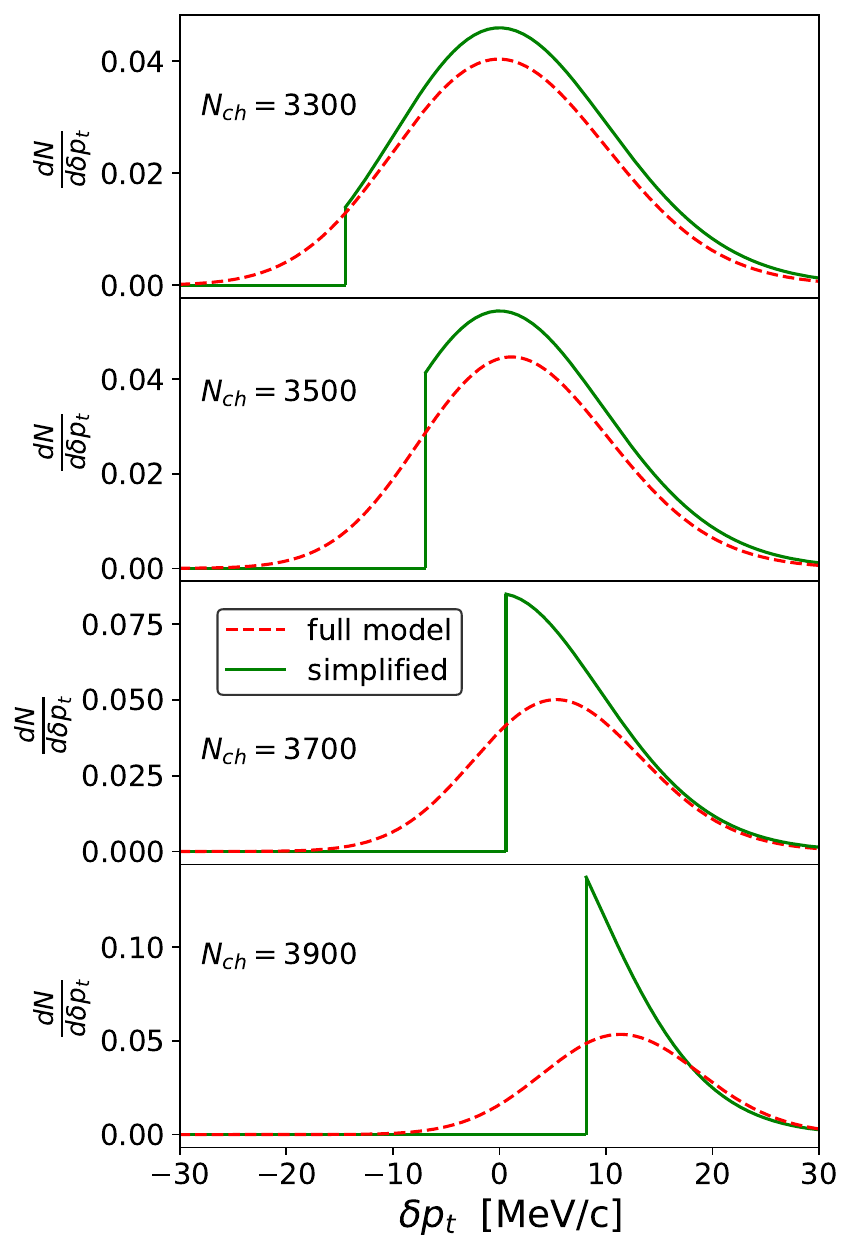} 
\end{center}
\caption{
Probability distribution of $[p_t]$ for various values of the multiplicity $N_{ch}$. 
If one defines centrality according to $N_{ch}$, the centrality fractions corresponding to these values of $N_{ch}$ are, from top to bottom, $2.2\%$, $1.1\%$, $0.3\%$ and $0.04\%$. 
We decompose $[p_t]$ as $[p_t]=p_{t0}+\delta p_t$, where $p_{t0}$ is a constant, and we only plot the distribution of  $\delta p_t$. 
The solid lines correspond to the simplified model where $[p_t]$ only depends on $N_{ch}$ and impact parameter (Eq.~(\ref{linearized})).
The dashed lines correspond to a more realistic model, assuming Gaussian fluctuations of $[p_t]$ for fixed $N_{ch}$ and $c_b$. 
}
\label{fig:ptdistribution}
\end{figure} 
The probability distribution of $c_b$ for fixed $N_{ch}$ is then given by Bayes' theorem:\footnote{$c_b$ is the cumulative distribution of $b$, therefore, $p(c_b)=1$ by construction.}
\begin{eqnarray}
  \label{gaussiancb}
  p(c_b|N_{ch})&=&\frac{p(N_{ch}|c_b)}{p(N_{ch})}\cr
&\propto& \exp\left(-\frac{\left(N_{ch}-N_{\rm knee}+\beta c_b\right)^2}{2\sigma_{N_{ch}}^2}\right),
\end{eqnarray}
where we have used Eqs.~(\ref{gaussianN}) and (\ref{linearizedmean}). Eq.~(\ref{gaussiancb}) shows that the distribution of $c_b$ is Gaussian, with a width $\sigma_{N_{ch}}/\beta\simeq 0.9\%$. 
It is, however, a {\it truncated\/} Gaussian, because of the boundary condition $c_b\ge 0$~\cite{Das:2017ned}.
Eq.~(\ref{linearized}) then implies that the probability distribution of $[p_t]$ is also a truncated Gaussian, with the boundary condition $[p_t]\ge p_t^{\rm min}$. 
This is illustrated by the solid curves in Fig.~\ref{fig:ptdistribution}, which will be discussed in more detail below. 

This truncation has several effects.
First, the distribution of $[p_t]$ becomes narrower, resulting in a decrease of the variance. 
This decrease has been seen by ATLAS~\cite{ATLAS:2022dov} and analyzed in Ref.~\cite{Samanta:2023amp}. 
Second, the truncation generates non-Gaussian features such as skewness and kurtosis, which are the focus of this paper. 

We now introduce a more realistic model of $[p_t]$ fluctuations, in order to take into account that $[p_t]$ can vary even if both $N_{ch}$ and $c_b$ are fixed.
We assume, following Ref.~\cite{Samanta:2023amp}, that the joint distribution of $N_{ch}$ and $[p_t]$ at fixed $c_b$, $p([p_t],N_{ch}|c_b)$, is a correlated Gaussian. 
The simplified model above corresponds to the limit where the correlation is maximal and there is a one-to-one correspondence between  $[p_t]$ and $N_{ch}$  at fixed $c_b$. 

The Gaussian ansatz can be justified in the following way. 
In a hydrodynamic model, fluctuations of $N_{ch}$ and $[p_t]$ both stem from fluctuations of the initial density profile. 
At fixed $b$, these density fluctuations originate from quantum fluctuations, either in the wave functions of incoming nuclei~\cite{PHOBOS:2006dbo,Miller:2007ri,Gelis:2010nm} or in the collision dynamics. 
At ultrarelativistic energies, causality implies that fluctuations in different locations in the transverse plane are independent. 
Therefore, fluctuations of  $N_{ch}$ and $[p_t]$ can be thought of resulting from a large number of independent contributions, and the central limit theorem implies that their fluctuations are approximately Gaussian. 

The two-dimensional Gaussian distribution has five parameters: 
The mean and standard deviation of $N_{ch}$ and of $[p_t]$, and the Pearson correlation coefficient $r$ between $N_{ch}$ and $[p_t]$. 
All these parameters may depend on $c_b$. 
We now explain how they are obtained.
We use experimental data when possible, and model calculations otherwise. 

\begin{figure*}[h]
\begin{center}
\includegraphics[width=.8\linewidth]{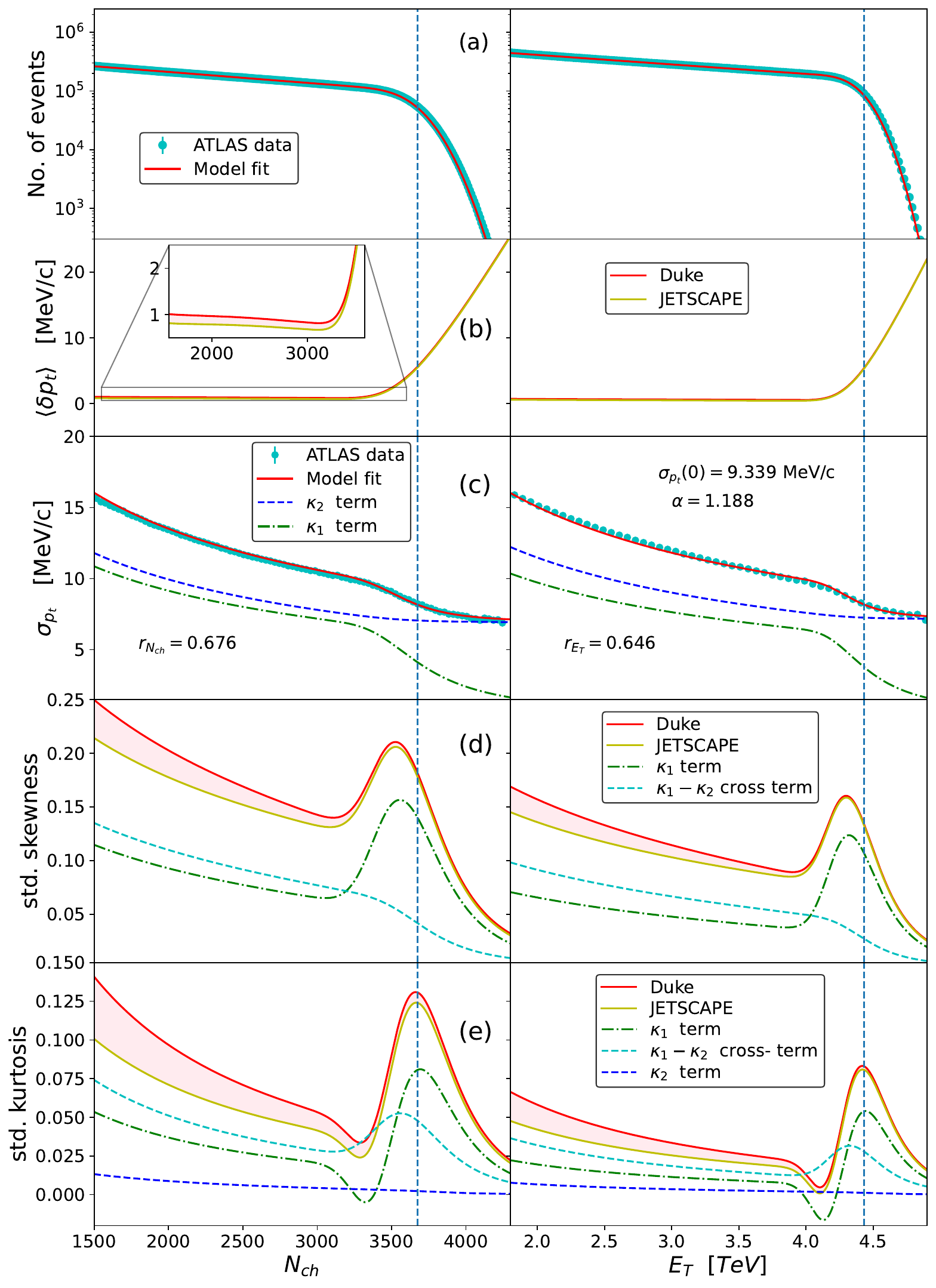} 
\end{center}
\caption{
(a)  Histogram of the number of charged particles $N_{ch}$ (left), measured in the inner detector of ATLAS, and of transverse energy $E_T$ (right), measured in the forward and backward calorimeters. 
The vertical dashed line indicates the position of the knee. 
The next four panels display the first cumulants of the distribution of $\delta p_t$, calculated using Eq.~(\ref{moments3b}), as a function of the centrality estimator.
(b) Mean. 
(c) Standard deviation ${\rm Var}(p_t)^{1/2}$. 
(d) Standardized skewness ${\rm Skew}(p_t)/{\rm Var}(p_t)^{3/2}$. 
(e) Standardized kurtosis ${\rm Kurt}(p_t)/{\rm Var}(p_t)^{2}$. 
The model is calibrated using ATLAS data for the standard deviation~\cite{ATLAS:2022dov}, shown as symbols in panel (c). 
In panels (b), (d) and (e), we display our predictions using the Duke and JETSCAPE parametrizations of the centrality dependence of $\sigma_{N_{ch}}$ (App.~\ref{s:trento}). 
The difference between the two is negligible for the mean (see zoom in panel (b)) and the standard deviation (not shown), but is sizable for the skewness and kurtosis, where it is displayed as a pink shaded band. 
The contributions of the various terms in Eq.~(\ref{skew}) are shown for the Duke parametrization only. 
}
\label{fig:panelplot}
\end{figure*}

From the probability distribution of the charged multiplicity $P(N_{ch})$, which is accurately measured (Fig.~\ref{fig:panelplot} (a)), one can infer the $c_b$-dependence of the mean multiplicity, $\overline{N_{ch}}(c_b)$, as well as the standard deviation $\sigma_{N_{ch}}$ for $c_b=0$~\cite{Das:2017ned,Samanta:2023amp}
On the other hand, the $c_b$ dependence of $\sigma_{N_{ch}}$ is not at all constrained by existing data. 
We therefore borrow this information from state-of-the-art models which have been tuned to experiment through Bayesian analyses. 
We use the Maximum A Posteriori parameter set from two analyses, one by the Duke group \cite{Moreland:2018gsh} and the other by the JETSCAPE collaboration (using the Grad viscous correction to the distribution function at particlization)~\cite{JETSCAPE:2020mzn}. 
The JETSCAPE analysis is tuned to a larger set of data, including several collision energies.
The Duke analysis is specifically tuned to 5.02~TeV data, which are the ones we use in this paper, and differs from the JETSCAPE analysis in the sense that nucleon substructure is taken into account, which may have an effect on fluctuations. 
We evaluate $\sigma_{N_{ch}}(c_b)$ for both models, in a way which is explained in detail in App.~\ref{s:trento}. 
The Duke parametrization predicts that $\sigma_{N_{ch}}$  increases between $b=0$ and $b=3.5$~fm, while the JETSCAPE parametrization predicts a slight decrease.
We use the difference between these two models as an estimate of the error in our predictions. 

The other parameters of our Gaussian model are the mean, $p_{t0}$, and the standard deviation, $\sigma_{p_t}$,  of $[p_t]$ at fixed impact parameter, and the Pearson correlation coefficient $r$ between $N_{ch}$ and $[p_t]$. 
Based on the observation that the variation of the mean value of $p_t$ is below the percent level in the 0-30\% centrality range~\cite{ALICE:2018hza}, we assume that $p_{t0}$ is independent of impact parameter. 
We decompose $[p_t]=p_{t0}+\delta p_t$, and we only model the distribution of $\delta p_t$, so that our results are independent of $p_{t0}$. 
We assume that $\sigma_{p_t}$ varies with $c_b$ like a power law of the mean multiplicity: 
\begin{equation}
\label{ptvariance}
\sigma_{p_t}(c_b)=\sigma_{p_t}(0)\left(\frac{N_{\rm knee}}{\overline{N_{ch}}(c_b)}\right)^{\alpha/2}, 
\end{equation} 
and we assume for simplicity that $r$ is independent of $c_b$. 
The three parameters $\sigma_{p_t}(0)$, $\alpha$, and $r$ are fitted to ATLAS data on the variance of $p_t$ fluctuations~\cite{ATLAS:2022dov,Samanta:2023amp}, in a range corresponding roughly to the 20\% most central collisions. 
These data imply in particular that $[p_t]$ is strongly correlated with the density, which is reflected in the correlation coefficient $r$, which is close to $0.7$. 

A property of the two-dimensional Gaussian is that if one fixes one of the variables, e.g. $N_{ch}$, the distribution of the other variable, e.g. $\delta p_t$, is Gaussian (this property will be used below in order to evaluate the skewness and kurtosis).  
The distribution of  $\delta p_t$ at fixed $N_{ch}$ and $c_b$ is defined by 
\begin{equation}
\label{fixedNandb}
p(\delta p_t|N_{ch},c_b)=\frac{p(\delta p_t,N_{ch}|c_b)}{p(N_{ch}|c_b)}. 
\end{equation} 
The distribution of $\delta p_t$ at fixed $N_{ch}$ is then obtained by averaging over impact parameter: 
\begin{eqnarray}
\label{cbintegral}
p(\delta p_t|N_{ch})&=&\int_0^1 p(\delta p_t|N_{ch},c_b)p(c_b|N_{ch})dc_b\cr
&=&\frac{1}{p(N_{ch})}\int_0^1 p(\delta p_t,N_{ch}|c_b)dc_b,
\end{eqnarray}
where we have used Eqs.~(\ref{gaussiancb}) and (\ref{fixedNandb}) in going from the first to the second line. 
The distributions $p(\delta p_t|N_{ch})$ are displayed as dashed lines in Fig.~\ref{fig:ptdistribution} for selected values of $N_{ch}$ near the knee. 
The full lines in this figure are obtained by setting the correlation coefficient to its maximum value $r=1$, corresponding to the simplified model of Eq.~(\ref{linearized}).  

Once the distribution of $\delta p_t$ is known, one can evaluate its cumulants. 
We derive semi-analytic expressions of the cumulants by using the property that the distribution of $\delta p_t$ is Gaussian for fixed $c_b$:
\begin{equation}
\label{1dgaussian}
P(\delta p_t|N_{ch},c_b)=\frac{1}{\sqrt{2\pi\kappa_2(c_b)}}\exp\left(-\frac{\left(\delta p_t-\kappa_1(c_b)\right)^2}{2\kappa_2(c_b)}\right),   
\end{equation}
where we omit the dependence on $N_{ch}$ in the right-hand side. 
$\kappa_1(c_b)$ and $\kappa_2(c_b)$ are the mean and the variance at fixed $N_{ch}$ {\it and\/} $c_b$, given by \cite{Samanta:2023amp}:
 \begin{eqnarray}
  \label{fixednch}
       \kappa_1(c_b)&=&r\frac{\sigma_{p_t}(c_b)}{\sigma_{N_{ch}}(c_b)}\left(N_{ch}-\overline{N_{ch}}(c_b)\right)\cr
       \kappa_2(c_b)&=& \left(1-r^2\right)\sigma_{p_t}^2(c_b).
\end{eqnarray}
Note that in the limit $r\to 1$, the variance $\kappa_2(c_b)$ vanishes and  $P(\delta p_t|N_{ch},c_b)$ reduces to a Dirac peak $\delta(\delta p_t-\kappa_1(c_b))$, implying that  $\delta p_t$ is solely determined by $N_{ch}$ and $c_b$. 
 
The moment of order $n$ is obtained by multiplying Eq.~(\ref{1dgaussian}) with $\delta p_t^n$ and integrating over $\delta p_t$. 
One thus obtains the following expressions for the first four moments: 
\begin{eqnarray}
  \label{moments3b}
  \langle \delta p_t|c_b\rangle&=&\kappa_1\cr
  \langle \delta p_t^2|c_b\rangle&=&\kappa_1^2+\kappa_2\cr
  \langle \delta p_t^3|c_b\rangle&=&\kappa_1^3+3\kappa_2\kappa_1\cr
  \langle \delta p_t^4|c_b\rangle&=&\kappa_1^4+6\kappa_2\kappa_1^2+3\kappa_2^2, 
\end{eqnarray}
where the dependence on $c_b$ on the right-hand side is implicit. 
These moments must then be averaged over $c_b$, as in Eq.~(\ref{cbintegral}). 
The cumulants are finally obtained from the moments using standard inversion formulas.  
The first four cumulants are the mean, the variance, the skewness and the excess kurtosis~\cite{Bhatta:2021qfk}: 
\begin{align}
  \label{skew}
  \langle\delta p_t\rangle&=\langle\kappa_1\rangle\cr
        {\rm Var}(p_t)&=(\langle\kappa_1^2\rangle-\langle\kappa_1\rangle^2)+\langle\kappa_2\rangle\cr
        {\rm Skew}(p_t)
        &=\langle\kappa_1^3\rangle-3\langle\kappa_1^2\rangle\langle\kappa_1\rangle+2\langle\kappa_1\rangle^3\cr
       &+3(\langle\kappa_2\kappa_1\rangle-\langle\kappa_2\rangle \langle\kappa_1\rangle)\cr
        {\rm Kurt}(p_t) &=\langle\kappa_1^4\rangle-4\langle\kappa_1^3\rangle\langle\kappa_1\rangle+6\langle\kappa_1^2\rangle\langle\kappa_1\rangle^2-3\langle\kappa_1\rangle^4\cr
       &+6(\langle\kappa_2\kappa_1^2\rangle-\langle\kappa_2\rangle \langle\kappa_1^2\rangle-2\langle\kappa_2\kappa_1\rangle\langle\kappa_1\rangle+2\langle\kappa_2\rangle\langle\kappa_1\rangle^2) \cr
       &+3(\langle\kappa_2^2\rangle-\langle\kappa_2\rangle^2). 
\end{align}
where angular brackets denote averages over $c_b$. 
One sees that $\kappa_1$ and $\kappa_2$  contribute separately to the variance. 
The term involving $\kappa_1$ is responsible for the sharp decrease of the variance around the knee, as shown in  Ref.~\cite{Samanta:2023amp}. 
The skewness has terms involving $\kappa_1$ only, and a term proportional to the correlation between $\kappa_1$ and $\kappa_2$. 
The kurtosis has one more term, which is proportional to the variance of $\kappa_2$. 
The skewness and the kurtosis encode the non-Gaussian properties of the event-by-event fluctuations of $[p_t]$. 
In our model, which assumes Gaussian fluctuations at fixed impact parameter, all non-Gaussianities originate from impact parameter fluctuations. 
If the impact parameter does not fluctuate, each line in the above expressions of ${\rm Skew}(p_t)$ and ${\rm Kurt}(p_t)$ is identically zero. 

Our quantitative predictions are displayed in panels (b), (d) and (e) of Fig.~\ref{fig:panelplot}.
The increase of the mean, displayed in panel (b), has already been discussed in the literature~\cite{Samanta:2023amp,Gardim:2019brr,Nijs:2021clz}. 
The new results of this paper are the skewness and the kurtosis, which both display sharp variations around the knee. 
We predict an increase of the skewness below the knee [such an increase has already been seen by the ALICE collaboration~\cite{Saha:2022bxf}, as will be discussed below], followed by a fast decrease above the knee. 
The kurtosis has first a minimum, followed by a maximum roughly at the knee. 
These structures come from the terms involving $\kappa_1$, and are inherited from the truncated Gaussian.
The cumulants of the truncated Gaussian (\ref{gaussiancb}) can be calculated analytically. 
The maximum of the skewness occurs at $N_{ch}\simeq N_{\rm knee}-\sigma_{N_{ch}}\simeq 3510$. 
The kurtosis has a minimum at $N_{ch}\simeq N_{\rm knee}-2\sigma_{N_{ch}}\simeq 3340$, followed by a maximum at  $N_{ch}\simeq N_{\rm knee}\simeq 3680$. 
This corresponds to the structure seen in our numerical results.

One sees that our predictions depend little on which scenario (Duke or JETSCAPE) one chooses for the centrality dependence of multiplicity fluctuations.
The main limitation of our model is that we have assumed a Gaussian distribution of $[p_t]$ at fixed $N_{ch}$ and $b$.
Since $[p_t]$ is a positive quantity, one expects its distribution to have a positive skewness $\kappa_3$ and a positive excess kurtosis $\kappa_4$.
This will give additional positive contributions to ${\rm Skew}(p_t)$ and ${\rm Kurt}(p_t)$ in Eq.~(\ref{skew}), of the form $\langle\kappa_3\rangle$ and $\langle\kappa_4\rangle$, so that our predictions should be considered a lower bound, both for the skewness and for the kurtosis.
Predicting quantitatively the value of this additional term is difficult and would require high-statistics hydrodynamic simulation.
We can however safely state that the additional contributions should have a smooth dependence on $N_{ch}$, and will typically result in a positive offset from our prediction. 
The sharp {\it variations\/} of the skewness and kurtosis around the knee in Fig.~\ref{fig:panelplot} (c) and (d) are robust, quantitative predictions. 

The ATLAS collaboration also carries out analyses by estimating the centrality using the transverse energy $E_T$ deposited in two calorimeters symmetrically with respect to the collision point, at smaller angles with respect to the beam than the inner detector. 
The whole analysis can be repeated by replacing $N_{ch}$ with $E_T$ everywhere, as shown in the right panel of Fig.~\ref{fig:panelplot}. 
It turns out that $E_T$ is a better centrality estimator than $N_{ch}$, which results in smaller impact parameter fluctuations~\cite{Pepin:2022jsd,Yousefnia:2021cup}. 
Since, in our model, all the non-Gaussianities originate from impact parameter fluctuations, one expects that both the skewness and the kurtosis are smaller if the centrality is determined as a function of $E_T$, which is exactly seen in our predictions. 
Experimental verification of these predictions will be crucial in assessing the importance of impact parameter fluctuations. 

Finally, let us comment on the recent preliminary results on the skewness released by the ALICE collaboration~\cite{Saha:2022bxf}.
The centrality estimator is the amplitude deposited in scintillators located at forward rapidities, qualitatively similar to the $E_T$-based centrality determination of ATLAS. 
The skewness is then determined in the central pseudorapidity ($\eta$) region, again similar to the ATLAS analysis, although with a narrower interval in $\eta$. 
The binning in centrality is much coarser than that of ATLAS, with each point corresponding to a $5\%$ interval. 
Our analysis covers roughly the $20\%$ most central collisions, therefore, our predictions can only be compared with the last four data points of ALICE, which correspond to the ranges (in TeV) $1.6<E_T<2.1$, $2.1<E_T<2.7$, $2.7<E_T<3.5$, and $E_T>3.5$. 
The quantity shown by ALICE is the intensive skewness~\cite{Giacalone:2020lbm}, not the standardized skewness. 
It is obtained by multiplying the standardized skewness with the mean, $\langle p_t\rangle$, and dividing by the standard deviation ${\rm Var}(p_t)^{1/2}$. 
We have not evaluated this quantity because the ATLAS collaboration does not provide the value of $\langle p_t\rangle$.  
Given the $p_t$ range covered by ATLAS, a rough guess is $\langle p_t\rangle\sim 1$~GeV$/c$. 
We then predict that the intensive skewness is essentially constant and close to $10$ in the interval $1.8<E_T<3.5$~TeV, while ALICE values increase from (slightly below) 4 to (slightly above) 5 in the equivalent range. 
However, one should not compare the absolute values, because they depend on the  $p_t$ coverage, which is $0.2<p_t<3$~GeV$/c$ for ALICE, and $0.5<p_t<5$~GeV$/c$ for ATLAS.\footnote{
The dependence of $\sigma_{p_t}(c_b)$ on the $p_t$ interval is not trivial. 
How the fluctuation of the $p_t$ spectrum depends on $p_t$ is at present not known, and assessing it would require to measure the quantity $v_0(p_t)$ recently introduced in Ref.~\cite{Schenke:2020uqq}.}
The interesting observation of ALICE is that the intensive skewness in the most central bin is close to $8$, significantly higher than in the previous bins. 
This last point of ALICE corresponds to the interval $E_T>3.5$~TeV, over which our calculation predicts a rise and fall of the intensive skewness, which peaks at a value $\sim 18$ below the knee. 
It is tempting to see in the ALICE result a first confirmation of our prediction. 
It will be useful if the ALICE analysis is repeated in finer centrality bins, and if they specify the value of the centrality estimator in each bin. 

\begin{acknowledgments}
We thank Somadutta Bhatta and Jiangyong Jia for suggesting us to publish quantitative predictions for the skewness, and Swati Saha and Bedangadas Mohanty for discussions about the ALICE preliminary results. 
We thank the Institute for Nuclear Theory at the University of Washington for hosting the program ``Intersection of nuclear structure and high-energy nuclear collisions'' during which this work was initiated. 
R. S. is supported by the Polish grant NAWA PRELUDIUM BIS: PPN/STA/2021/1/00040/U/00001 and the NCN grant PRELUDIUM BIS: 2019/35/O/ST2/00357.
M. L. thanks the S\~ao Paulo Research Foundation (FAPESP) for support under grants 2021/08465-9, 2018/24720-6, and 2017/05685-2, as well as the support of the Brazilian National Council for Scientific and Technological Development (CNPq).
We acknowledge support from the ``Emilie du Ch\^atelet'' visitor programme and from the GLUODYNAMICS project funded by the ``P2IO LabEx (ANR-10-LABX-0038)'' in the framework ``Investissements d'Avenir'' (ANR-11-IDEX-0003-01) managed by the Agence Nationale de la Recherche (ANR).
\end{acknowledgments}

\appendix
\section{Centrality dependence of multiplicity fluctuations}
\label{s:trento}

\begin{figure}[h]
\begin{center}
\includegraphics[width=\linewidth]{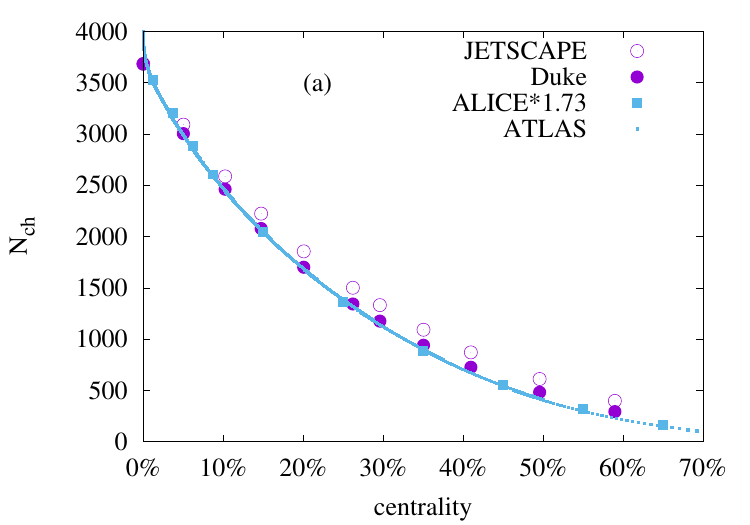} 
\includegraphics[width=\linewidth]{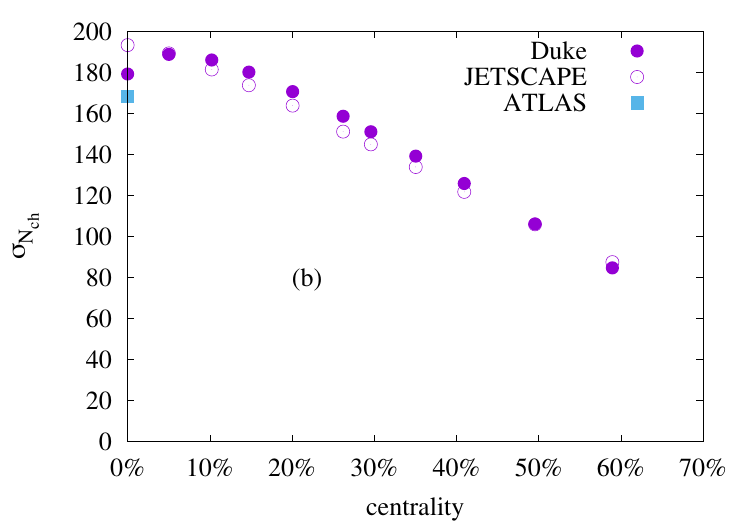} 
\end{center}
\caption{
(a)  Variation of charged multiplicity $N_{ch}$ with centrality in Pb+Pb collisions at $\sqrt{s_{NN}}=5.02$~TeV measured by ATLAS~\cite{ATLAS:2022dov} and ALICE~\cite{ALICE:2015juo}.
  For ATLAS, the centrality is defined from the cumulative distribution of $N_{ch}$ and then divided by a calibration factor $1.153$~\cite{ATLAS:2018ezv},which corrects for the fact that for the largest centrality fractions, some of the recorded events are fake.  
  The ALICE results have been rescaled by a factor $1.73$ to correct for the different acceptance and efficiency of the detector.
  The circles display the centrality dependence of the mean initial energy for the T$_{\text{R}}$ENTo parametrizations used by the Duke~\cite{Moreland:2018gsh} and JETSCAPE analyses~\cite{JETSCAPE:2020mzn}. 
The centrality is defined as $\pi b^2/\sigma_{\rm Pb}$, where $\sigma_{\rm Pb}=767$~fm$^2$ is the total inelastic cross section. 
(b) Variation of the standard deviation of $N_{ch}$ with centrality. 
}
\label{fig:mean}
\end{figure} 

The probability distribution of the multiplicity at fixed impact parameter $b$ is expected to be approximately Gaussian~\cite{Das:2017ned} and can be characterized by its mean $\overline{N_{ch}}$ and standard deviation $\sigma_{N_{ch}}$, which both depend on $b$. 
The mean can be reconstructed using the simple following rule. 
If a fraction $c_b$ of events have a multiplicity larger than $N$, then $N\simeq \overline{N_{ch}} (c_b)$~\cite{Broniowski:2001ei}. 
This simple rule, which is applied to ATLAS data in Fig.~\ref{fig:mean} (a), works well except for multiplicities around and above the knee. 

On the other hand, the centrality dependence of $\sigma_{N_{ch}}$ is not known, and we use state-of-the-art hydrodynamic calculations by the Duke group~\cite{Moreland:2018gsh} and by the JETSCAPE collaboration~\cite{JETSCAPE:2020mzn} to evaluate it. 
However, we want to avoid running massive hydrodynamic calculations, and we therefore  estimate the multiplicity fluctuations from the initial conditions of these calculations. 
We assume that for every collision event, the multiplicity is proportional to the initial energy.
Both Duke and JETSCAPE analyses employ the T$_{\text{R}}$ENTo parametrization~\cite{Moreland:2014oya} for the initial energy density, but with slightly different values of the parameters. 
We run these T$_{\text{R}}$ENTo initial conditions for several fixed values of $b$ (specifically, $b=0$, $3.5$, $5$, $6$, $7$, $8$, $8.5$, $9.25$, $10$, $11$, $12$~fm). 
For each $b$, we generate $10^5$ events with both Duke and JETSCAPE parameters, and we compute the initial energy of each event. 
We rescale this energy by a constant factor so that it matches the ATLAS result for the charged multiplicity at $b=0$~\cite{Samanta:2023amp}.
The variation of the mean $N_{ch}$ with centrality is displayed in Fig.~\ref{fig:mean} (a).
Experimental data are also shown. 
One sees that ALICE and ATLAS data are in excellent agreement once properly rescaled.
The calculation using the Duke parametrization agrees very well with experiment. 
Agreement is not quite as good, but still reasonable, for the JETSCAPE parametrization. 

We then calculate the standard deviation of $N_{ch}$, $\sigma_{N_{ch}}$, for each value of $b$. 
Results are displayed in Fig.~\ref{fig:mean} (b). 
The standard deviation can only be measured at $b=0$~\cite{Yousefnia:2021cup} from the tail of the distribution of $N_{ch}$, therefore, there is only one data point on this plot.
One sees that both model calculations are in reasonable agreement with this data point, but slightly overestimate it. 

\begin{figure}[h]
\begin{center}
\includegraphics[width=\linewidth]{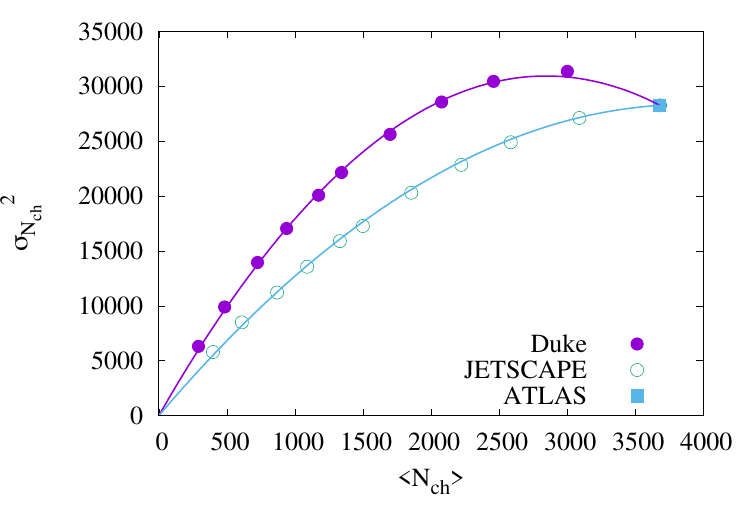} 
\end{center}
\caption{Parametric plot of the mean and variance of $N_{ch}$ as a function of $b$.
  Both models have been calibrated in such a way that they match data at $b=0$.
  Solid lines are fits using $y=\gamma x+(1-\gamma) x^2$, where $y\equiv\sigma^2_{N_{ch}}(b)/\sigma^2_{N_{ch}}(0)$ and $x\equiv \langle N_{ch}\rangle(b)/\langle N_{ch}\rangle(0)$, with $\gamma=2.83$ (Duke) and $\gamma=1.90$ (JETSCAPE). 
The calculation in Ref.~\cite{Samanta:2023amp} was done with $\gamma=1$ (variance proportional to mean). 
}
\label{fig:variance}
\end{figure} 

We use model calculations only to predict the $b$-dependence of $\sigma_{N_{ch}}$, not the value at $b=0$ which is measured precisely. 
We therefore rescale $\sigma_{N_{ch}}$ from the model calculation by a constant factor so that it matches the experimental value at $b=0$. 
The resulting predictions for $b>0$ are displayed in Fig.~\ref{fig:variance}.
We plot the variance $\sigma_{N_{ch}}^2$ as a function of the mean.
If $N_{ch}$ is the sum of $k$ identical, uncorrelated distributions, where $k$ depends on $b$, both the mean and the variance are proportional to $k$, therefore, they are proportional to one another.
This behavior is only observed for large values of $b$.
Both model calculations predict that the variance increases more slowly as $b$ decreases. 
The Duke calculation even predicts that it decreases for the smallest values of $b$. 
The two solid lines, which are polynomial fits to our calculations, are used as two limiting cases which define the error bands in Fig.~\ref{fig:panelplot}.

\end{document}